\documentclass[journal]{IEEEtran}
\ifCLASSINFOpdf
\else
\fi
\begin{document}

\title{Necessary and sufficient conditions for positive semidefinite quantum mutual information matrices}

\author{Feng Liu, Fei Gao, Su-Juan Qin, and Qiao-Yan Wen
\thanks{This work is supported by NSFC (Grant Nos. 61300181, 61272057, 61202434, 61170270, 61100203, 61121061), Beijing Natural Science Foundation (Grant No. 4122054), Beijing Higher Education Young Elite Teacher Project (Grant Nos. YETP0475, YETP0477), and BUPT Excellent Ph.D. Students Foundation (Grant No. CX201434).

The authors are with State Key Laboratory of Networking and Switching Technology, Beijing University of Posts and Telecommunications, Beijing, 100876, China. Feng Liu is also with School of Mathematics and Statistics Science, Ludong University, Yantai 264025, China. (e-mail: gaofei\_bupt@hotmail.com).}
\thanks{}}

%
%

\markboth{}%
{}
\maketitle

\begin{abstract}
For any $n$-partite state $\rho_{A_{1}A_{2}\cdot\cdot\cdot A_{n}}$, we define its quantum mutual information matrix as an $n$ by $n$ matrix whose $(i,j)$-entry is given by quantum mutual information $I(\rho_{A_{i}A_{j}})$. Although
each entry of quantum mutual information matrix, like its classical counterpart, is also used to measure bipartite correlations, the similarity ends here: quantum mutual information matrices are not always positive semidefinite even for collections of up to 3-partite states. In this work, we obtain necessary and sufficient conditions for the positive semidefinite quantum mutual information matrix. We further define the \emph{genuine} $n$-partite mutual information which can be easily calculated. This definition is symmetric, nonnegative, bounded and more accurate for measuring multipartite states.
\end{abstract}


%
\IEEEpeerreviewmaketitle

\section{Introduction}

In classical information theory, the Shannon entropy $H(X)=H(p)=-\sum_{i}p_{i}\log_{2}p_{i}$ is used to
quantify the information in a source $X$, which produces messages $x_{i}$ with a probability distribution $p=\{p_{i}\}$. Correlations between two different discrete random variables $X$ and $Y$ are measured by the mutual information

~~~~~~~~~~$I(X:Y)=H(X)+H(Y)-H(XY)$.\\
The mutual information measures how much information $X$ and $Y$ have in common, and it is bounded above by the marginal entropies:
\begin{equation}
I(X:Y)\leq min\{H(X),H(Y)\}.
\end{equation}

By analogy with the mutual information it is possible to define quantum mutual information for composite quantum systems. When $\rho_{A_{1}A_{2}}$ is shared by two parties $A_{1}$ and $A_{2}$ with marginals $\rho_{A_{1}}=tr_{A_{2}}\rho_{A_{1}A_{2}}$ and $\rho_{A_{2}}=tr_{A_{1}}\rho_{A_{1}A_{2}}$, the straightforward generalization of the mutual information is the quantum mutual information
\begin{equation}
I(\rho_{A_{1}A_{2}})=S(\rho_{A_{1}})+S(\rho_{A_{2}})-S(\rho_{A_{1}A_{2}}), \end{equation}
where $S(\rho_{A_{i}})$ is the von Neumann entropy. The quantum mutual information is also used to quantify the total correlations in $\rho_{A_{1}A_{2}}$ [1,2]. As a rule of thumb, the quantum world is full of surprises as pointed out by Li and Luo in Ref. [3], and indeed new phenomena arise here. (i) Suppose $|A_{1}A_{2}\rangle$ is a pure state, $S(A_{1}|A_{2})<0$ if and only if $|A_{1}A_{2}\rangle$ is entangled. In other words, the conditional entropy can be either positive, negative, or zero in a general composite quantum system. (ii) Based on quantum
effects, which lead to stronger correlations than classically
possible, one has the following bound:
\begin{equation}
I(\rho_{A_{1}A_{2}})\leq 2min\{S(\rho_{A_{1}}),S(\rho_{A_{2}})\}.
\end{equation}
The factor 2 is apparently of a quantum origin [3]. In particular, if $S(A_{1}|A_{2})<0$, then

~~~~~~~~$I(\rho_{A_{1}A_{2}})=S(A_{1})-S(A_{1}|A_{2})>S(A_{1})$.\\ This phenomenon has many interesting applications for quantum information theory.

In order to effectively registration of multiple ultrasound images, Wang and Shen [4] introduce the mutual information matrix. This matrix has been conjectured to be positive semidefinite. Recently, Jakobsen [5] gave counterexamples to the conjecture, and shew that the conjecture holds for up to three random variables. By analogy with the mutual information matrix we give the definition for quantum mutual information matrix. Motivated by the Jakobsen's result [5], one might be tempted to guess that the quantum mutual information matrix is always positive semidefinite for up to a 3-partite state. Amazingly, due to quantum effects (i) and (ii), which lead to stronger correlations than classically possible, the above conjecture does not hold in general, and then the necessary and sufficient conditions should be considered. This is the question we address in this work. On the other hand, Polani [5] has observed that the mutual information matrix is positive semidefinite in many applications. How to give a naturally general sufficient condition that explains this phenomenon is an open problem. Exploiting the eigenvalues of the contract diagonal matrix with quantum mutual information matrices and the form of the Shannon entropies, we define the \emph{genuine} $n$-partite quantum
mutual information, and discuss its properties. Then, the definition is testified in multi-qubit pure states and proved to be more accurate than $n$-partite information [6,7] on measurement the total correlation of multipartite states.

In the next section we introduce the quantum mutual information matrix and give a counterexample to show that it is not always positive semidefinite. Section III contains our proof that the necessary and sufficient conditions for the positive semidefinite mutual information matrix. We also explain why we believe that these conditions can be continually hold when the number of partite is increasing. In Sec. IV, we describe the \emph{genuine} $n$-partite quantum mutual information, and show that it is more effective on measurement of the total correlation. We conclude in Sec. V.
\section{Quantum mutual information matrix}
For any $n$-partite state $\rho_{A_{1}A_{2}\cdot\cdot\cdot A_{n}}$, we define its quantum mutual information matrix to be the $n$ by $n$ matrix whose $(i,j)$ entry is given by $I(\rho_{A_{i}A_{j}})$ where

~~~~~~~~$\rho_{A_{i}A_{j}}=tr_{\{A_{1}A_{2}\cdot\cdot\cdot A_{n}\}-\{A_{i}A_{j}\}}\rho_{A_{1}A_{2}\cdot\cdot\cdot A_{n}}$.\\ In particular,

~~~~~~$I(\rho_{A_{i}A_{i}})=S(\rho_{A_{i}})$, $I(\rho_{A_{i}A_{j}})=I(\rho_{A_{j}A_{i}})$ and

~~~~~~~~~~~~~~~~~~~~~$I(\rho_{A_{i}A_{j}})\geq 0$.\\
Then the quantum mutual information matrix is as follows:
\begin{equation}       
M_{n}=\left(                 
  \begin{array}{cccc}   
    S(\rho_{A_{1}}) & I(\rho_{A_{1}A_{2}}) & \cdots & I(\rho_{A_{1}A_{n}})\\  
    I(\rho_{A_{1}A_{2}}) & S(\rho_{A_{2}}) & \cdots & I(\rho_{A_{2}A_{n}})\\
    \vdots & \vdots & \vdots & \vdots\\
    I(\rho_{A_{1}A_{n}}) & I(\rho_{A_{2}A_{n}}) & \cdots & S(\rho_{A_{n}})\\
  \end{array}
\right)                 
\end{equation}
which is obviously a real symmetric matrix. Then the quantum mutual information matrix and its contract matrix $M'_{n}$ ($M'_{n}=C^{T}M'_{n}C$ where $C$ is an invertible matrix, and  $C^{T}$ is the matrix transpose of $C$) have the same index of inertia.

In the classical world, the mutual information matrix was proofed to be positive semi-definite for all three-tuples in [5]. The proof can be obtained from Eq.(1). Here, we will give a counterexample for 2 by 2 quantum mutual information matrix which is based on the negative conditional entropy.

\emph{Example 1}. Consider a system $A_{1}A_{2}$ of two qubits in the entangled state $\rho_{A_{1}A_{2}}=(|01\rangle-|10\rangle)/\sqrt{2}$. System $A_{1}$($A_{2}$) has the density operator $I/2$, and thus has entropy equal to one. On the other hand, this is a pure state so

~~~~~~$S(\rho_{A_{1}A_{2}})=0$ and $I(\rho_{A_{1}A_{2}})=2S(\rho_{A_{1}})=2$.\\
The quantum mutual information matrix for $\rho_{A_{1}A_{2}}$ is

~~~~~~~~~~~~~~~~~~~~$       
M_{2}=\left(                 
  \begin{array}{cc}   
    1 & 2 \\  
    2 & 1 \\
  \end{array}
\right)                 
$\\
which is contract with

~~~~~~~~~~~~~~~~~~~~$       
M'_{2}=\left(                 
  \begin{array}{cc}   
    1 & 0 \\  
    0 & -3 \\
  \end{array}
\right),                
$\\
i.e. there exists the invertible matrix
$       
F=\left(                 
  \begin{array}{cc}   
    1 & 0 \\  
    -2 & 1 \\
  \end{array}
\right),                
$
which transforms $M_{2}$ into $M'_{2}$ as follows $FM_{2}F^{T}=M'_{2}$. Because a matrix is positive semidefinite if and only if its eigenvalues are all nonnegative, $M_{2}$ is not a positive definite matrix.

The counterexample suggests that the negative conditional entropy [8] or $I(\rho_{A_{i}A_{j}})\geq S(\rho_{A_{i}})$ is responsible for this counterintuitive phenomenon. In the next section we study this phenomenon from a different perspective and find the necessary and sufficient conditions for positive semidefinite mutual information matrices.
\section{Necessary and sufficient conditions}

For any $n$-partite state $\rho_{A_{1}A_{2}\cdot\cdot\cdot A_{n}}$, we are going to determine which states' quantum mutual information matrices are positive semidefinite. Combining the inequality relation (3) with the zero entropy for a pure state, we have two lemmas as follows.

\emph{Lemma 1}. For the $n$-partite state $\rho_{A_{1}A_{2}\cdot\cdot\cdot A_{n}}$, if there exists $i\in \{1,2,\cdots,n\}$ satisfying $S(\rho_{A_{i}})=0$, then $I(\rho_{A_{i}A_{j}})=0$ for every $j\in \{1,2,\cdots,n\}$.

These equalities can be proved by Eq.(3) for any $2$-partite state $\rho_{A_{i}A_{j}}$. From Eq. (3), we can see that

~~~~~~~~~~~~~~~~~~$I(\rho_{A_{i}A_{j}})\leq 2S(\rho_{A_{i}})=0$. \\
Since $0\leq I(\rho_{A_{i}A_{j}})$, we know that $I(\rho_{A_{i}A_{j}})=0$ for every $j\in \{1,2,\cdots,n\}$.

From \emph{Lemma 1}, all elements in the i\emph{th} row and i\emph{th} column of the quantum mutual information matrix $M_{n}$ of $\rho_{A_{1}A_{2}\cdot\cdot\cdot A_{n}}$ are zero when $S(\rho_{A_{i}})=0$. Therefore, $M_{n}$ has the same positive semidefinite property with $M_{n-1}$ of $\rho_{A_{1}A_{2}\cdot\cdot\cdot A_{i-1}A_{i+1}\cdots A_{n}}$. This is our second Lemma.

\emph{Lemma 2}. For the $n$-partite state $\rho_{A_{1}A_{2}\cdot\cdot\cdot A_{n}}$, if there exists $i\in \{1,2,\cdots,n\}$ satisfying $S(\rho_{A_{i}})=0$, then $M_{n}$ and $M_{n-1}$ of $\rho_{A_{1}A_{2}\cdot\cdot\cdot A_{i-1}A_{i+1}\cdots A_{n}}$ have the same positive semi-definite property.

Since $M_{n}$ and its contract matrix have the same positive semi-definite property, we can also make $S(\rho_{A_{i}})$ in ascending order, i.e. $S(\rho_{A_{1}})\leq S(\rho_{A_{2}})\leq \cdots S(\rho_{A_{n}})$. Without loss of generality, hereafter the matrix always have this characteristic. Let the $i$th-order  principle minor sequence of $M_{n}$ is

~~~~$P_{i}=\left(                 
  \begin{array}{cccc}   
    S(\rho_{A_{1}}) & I(\rho_{A_{1}A_{2}}) & \cdots & I(\rho_{A_{1}A_{i}})\\  
    I(\rho_{A_{1}A_{2}}) & S(\rho_{A_{2}}) & \cdots & I(\rho_{A_{2}A_{i}})\\
    \vdots & \vdots & \vdots & \vdots\\
    I(\rho_{A_{1}A_{i}}) & I(\rho_{A_{2}A_{i}}) & \cdots & S(\rho_{A_{i}})\\
  \end{array}
\right).$\\
From the above lemmas, we give the explicit properties of the quantum states which saturate the positive semi-definite property of mutual information matrix, and have the following results.

\emph{Theorem 1}. For any $2$-partite state $\rho_{A_{1}A_{2}}$, we have

a) $M_{2}$ must be positive semi-definite, when

~~~~~~~~~~~~~~~$S(\rho_{A_{1}})\cdot S(\rho_{A_{2}})=0$.

b) $M_{2}$ is positive semi-definite if and only if

~~~~~~~~~~~$P_{2}=S(\rho_{A_{1}})\cdot S(\rho_{A_{2}})- I^{2}(\rho_{A_{1}A_{2}})\geq 0$, \\
when $P_{1}=S(\rho_{A_{1}})> 0$.

\emph{Proof}. For $\rho_{A_{1}A_{2}}$,

~~~~~~~~~~$       
M_{2}=\left(                 
  \begin{array}{cc}   
    S(\rho_{A_{1}}) & I(\rho_{A_{1}A_{2}}) \\  
    I(\rho_{A_{1}A_{2}}) & S(\rho_{A_{2}}) \\
  \end{array}
\right).                 
$

a) If there exists $i\in \{1,2\}$ satisfying $S(\rho_{A_{i}})=0$ and without loss of generality, let $S(\rho_{A_{1}})=0$, $M_{2}$ and $M_{1}=(S(\rho_{A_{2}}))$ have the same positive semi-definite property from \emph{Lemma 1} and \emph{Lemma 2}. $M_{1}$ is positive semi-definite if and only if $S(\rho_{A_{2}})\geq 0$, which is always stand up. Therefore, we obtain that

~~~~~~~~~~$S(\rho_{A_{1}})\cdot S(\rho_{A_{2}})= I^{2}(\rho_{A_{1}A_{2}})=0$ \\
and $M_{2}$ is positive semi-definite.

b) If $S(\rho_{A_{1}})>0$, $M_{2}$ is  contract with

~~~~~~~~~~$       
M'_{2}=\left(                 
  \begin{array}{cc}   
    S(\rho_{A_{1}}) & 0 \\  
    0 & \frac{S(\rho_{A_{1}})\cdot S(\rho_{A_{2}})- I^{2}(\rho_{A_{1}A_{2}})}{S(\rho_{A_{1}})} \\
  \end{array}
\right).                 
$

$M'_{2}$ is positive semi-definite if and only if its every diagonal elements are nonnegative, that is to say,

~~~~~~~~~~~$P_{2}=S(\rho_{A_{1}})\cdot S(\rho_{A_{2}})- I^{2}(\rho_{A_{1}A_{2}})\geq 0$. \\
Therefore, $M_{2}$ is positive semi-definite under the same condition.

This completes the proof of Theorem 1.

This result shows that $M_{2}$ is positive semi-definite if and only if its second order principle minor sequence is nonnegative. That is to say, Theorem 1 can be equivalently expressed as follows.

\emph{Theorem 1'}. For any $2$-partite state $\rho_{A_{1}A_{2}}$, $M_{2}$ is positive semi-definite if and only if all second order principle minor sequences are nonnegative, i.e. its determinant $|M_{2}|$ is nonnegative.

The phenomena is completely different from mutual information matrices in the classical world, because $H(X)\cdot H(Y)\geq I^{2}(X:Y)$ is always stand up for any two random variables $X$ and $Y$ from Eq.(1).

By analogy with the above processing of proof, it is possible to obtain the similar necessary and sufficient conditions for $3$-partite state as follows.

\emph{Theorem 2}. For any $3$-partite state $\rho_{A_{1}A_{2}A_{3}}$, under the concept of contract, we have

a) $M_{3}$ is positive semi-definite if and only if

~~~~$S(\rho_{A_{2}})\cdot S(\rho_{A_{3}})\geq I^{2}(\rho_{A_{2}A_{3}})$, when $S(\rho_{A_{1}})=0$.

b) $M_{3}$ is positive semi-definite if and only if

~~~~~~~~~~$S(\rho_{A_{1}})\cdot S(\rho_{A_{3}})\geq I^{2}(\rho_{A_{1}A_{3}})$

~~~~and $S(\rho_{A_{1}})\cdot I(\rho_{A_{2}A_{3}})=I(\rho_{A_{1}A_{2}})\cdot I(\rho_{A_{1}A_{3}})$, \\
when $S(\rho_{A_{1}})\geq 0$ and $S(\rho_{A_{1}})\cdot S(\rho_{A_{2}})= I^{2}(\rho_{A_{1}A_{2}})$.

c) $M_{3}$ is positive semi-definite if and only if\\
$S(\rho_{A_{1}})\cdot S(\rho_{A_{2}})\cdot S(\rho_{A_{3}})+2\cdot I(\rho_{A_{1}A_{2}})\cdot I(\rho_{A_{1}A_{3}})\cdot I(\rho_{A_{2}A_{3}})\geq S(\rho_{A_{1}})\cdot I^{2}(\rho_{A_{2}A_{3}})+S(\rho_{A_{2}})\cdot I^{2}(\rho_{A_{1}A_{3}})+S(\rho_{A_{3}})\cdot I^{2}(\rho_{A_{1}A_{2}})$,\\
i.e. $P_{3}\geq 0$,
when $P_{2}=S(\rho_{A_{1}})\cdot S(\rho_{A_{2}})- I^{2}(\rho_{A_{1}A_{2}})> 0$.

\emph{Proof}. For $\rho_{A_{1}A_{2}A_{3}}$,

~~~~~~~~~~$       
M_{3}=\left(                 
  \begin{array}{ccc}   
    S(\rho_{A_{1}}) & I(\rho_{A_{1}A_{2}})& I(\rho_{A_{1}A_{3}}) \\  
    I(\rho_{A_{1}A_{2}}) & S(\rho_{A_{2}}) & I(\rho_{A_{2}A_{3}})\\
    I(\rho_{A_{1}A_{3}}) & I(\rho_{A_{2}A_{3}}) & S(\rho_{A_{3}})\\
  \end{array}
\right).                 
$

a) If $S(\rho_{A_{1}})=0$, $M_{3}$ and

~~~~~~~~~~$       
M_{2}=\left(                 
  \begin{array}{cc}   
    S(\rho_{A_{2}}) & I(\rho_{A_{2}A_{3}}) \\  
    I(\rho_{A_{2}A_{3}}) & S(\rho_{A_{3}}) \\
  \end{array}
\right)                 
$\\
have the same positive semi-definite property from Lemma 2. From Theorem 1, we know $M_{2}$ is positive semi-definite if and only if $S(\rho_{A_{2}})\cdot S(\rho_{A_{3}})\geq I^{2}(\rho_{A_{2}A_{3}})$. So $M_{3}$ is positive semi-definite under the same limitation.

b) Because $S(\rho_{A_{1}})\neq0$, $M_{3}$ is contract with

~~~~~~~~~~$       
M'_{3}=\left(                 
  \begin{array}{ccc}   
    S(\rho_{A_{1}}) & 0& 0 \\  
    0 & \alpha & \beta\\
    0 & \beta & \gamma\\
  \end{array}
\right).                 
$\\
where
$\alpha=\frac{S(\rho_{A_{1}})\cdot S(\rho_{A_{2}})- I^{2}(\rho_{A_{1}A_{2}})}{S(\rho_{A_{1}})}$,

~~~~~~$\beta=\frac{S(\rho_{A_{1}})\cdot I(\rho_{A_{2}A_{3}})- I(\rho_{A_{1}A_{2}})\cdot I(\rho_{A_{1}A_{3}})}{S(\rho_{A_{1}})}$, and

~~~~~~$\gamma=\frac{S(\rho_{A_{1}})\cdot S(\rho_{A_{3}})- I^{2}(\rho_{A_{1}A_{3}})}{S(\rho_{A_{1}})}$.

When $S(\rho_{A_{1}})\cdot S(\rho_{A_{2}})- I^{2}(\rho_{A_{1}A_{2}})=0$, the necessary condition for positive semi-definite $M_{3}$ is its principle minor sequences are all nonnegative. Then we have

$\alpha\geq0$, $\gamma\geq0$, and
$\left|                 
  \begin{array}{cc}   
    \alpha & \beta \\  
    \beta & \gamma\\
  \end{array}
\right|$$\geq 0$.

However, $\left|                 
  \begin{array}{cc}   
    \alpha & \beta \\  
    \beta & \gamma\\
  \end{array}
\right|$$=-\beta^{2}\geq 0$ if and only if $\beta=0$. $M'_{3}$ can be rewritten as follows

~~~~~~~~~~$       
\left(                 
  \begin{array}{ccc}   
    S(\rho_{A_{1}}) & 0& 0 \\  
    0 & 0 & 0\\
    0 & 0 & \frac{S(\rho_{A_{1}})\cdot S(\rho_{A_{3}})- I^{2}(\rho_{A_{1}A_{3}})}{S(\rho_{A_{1}})}\\
  \end{array}
\right).                 
$

So we obtain that $M_{3}$ is positive semi-definite if and only if

$S(\rho_{A_{1}})\cdot S(\rho_{A_{3}})\geq I^{2}(\rho_{A_{1}A_{3}})$ and

$S(\rho_{A_{1}})\cdot I(\rho_{A_{2}A_{3}})=I(\rho_{A_{1}A_{2}})\cdot I(\rho_{A_{1}A_{3}})$.

c) When $S(\rho_{A_{1}})\cdot S(\rho_{A_{2}})- I^{2}(\rho_{A_{1}A_{2}})>0$, $M_{3}$ is contract with
\begin{equation}
M''_{3}=\left(                 
  \begin{array}{ccc}   
    S(\rho_{A_{1}}) & 0& 0 \\  
    0 & \xi & 0\\
    0 & 0 & \zeta\\
  \end{array}
\right).                 
\end{equation} \\
where
$\xi=\frac{S(\rho_{A_{1}})\cdot S(\rho_{A_{2}})- I^{2}(\rho_{A_{1}A_{2}})}{S(\rho_{A_{1}})}$, and

~~~~~~$\zeta=\frac{S(\rho_{A_{1}})\cdot S(\rho_{A_{3}})- I^{2}(\rho_{A_{1}A_{3}})}{S(\rho_{A_{1}})\cdot (S(\rho_{A_{1}})\cdot S(\rho_{A_{2}})- I^{2}(\rho_{A_{1}A_{2}}))}$

~~~~~~~~$-\frac{ (S(\rho_{A_{1}})\cdot I(\rho_{A_{2}A_{3}})- I(\rho_{A_{1}A_{2}})\cdot I(\rho_{A_{1}A_{3}}))^{2}}{S(\rho_{A_{1}})\cdot (S(\rho_{A_{1}})\cdot S(\rho_{A_{2}})- I^{2}(\rho_{A_{1}A_{2}}))^{2}}$.

So, $M_{3}$ is positive semi-definite if and only if $\zeta\geq 0$, i.e.\\
$S(\rho_{A_{1}})\cdot S(\rho_{A_{2}})\cdot S(\rho_{A_{3}})+2\cdot I(\rho_{A_{1}A_{2}})\cdot I(\rho_{A_{1}A_{3}})\cdot I(\rho_{A_{2}A_{3}})\geq S(\rho_{A_{1}})\cdot I^{2}(\rho_{A_{2}A_{3}})+S(\rho_{A_{2}})\cdot I^{2}(\rho_{A_{1}A_{3}})+S(\rho_{A_{3}})\cdot I^{2}(\rho_{A_{1}A_{2}})$.

This completes the proof of Theorem 2.

This result similarly shows that $M_{3}$ is positive semi-definite if and only if its every principle minor sequence are all nonnegative. That is to say, Theorem 2 can be equivalently expressed as follows.

\emph{Theorem 2'}. For any $3$-partite state $\rho_{A_{1}A_{2}A_{3}}$, $M_{3}$ is positive semi-definite if and only $P_{3}=|M_{3}|\geq 0$ when $P_{2}>0$.

The phenomena is also completely different from mutual information matrices in the classical world, because $M_{3}$ is always positive semi-definite for any three random variables [5]. In the similar way, we convince that $M_{n}$ is positive semi-definite if and only if its determinant $P_{n}=|M_{n}|$ is nonnegative when the $(n-1)$th-order principle minor sequence is positive.

In the next section, we will show that $M_{n}$ can be used to define the \emph{genuine} $n$-partite quantum mutual information, which supports Polani's assertion [5].
\section{Genuine quantum mutual information}

Quantum mutual information measures the total amount of correlation (both classical and quantum) between two systems. Consistent with this interpretation, the mutual information is always non-negative. To measures the total amount of correlation of $n$-partite state, the more complicated $n$-partite information [6] (or $I_{n}$-measure [7]) is defined. However, in a general quantum system $I_{n}$ can be either positive, negative, or zero, and there exists a typical quantum field theory which can exhibit all three behaviors depending on the choice of $n$ systems [6,9]. The phenomenon means that unlike quantum mutual information on $2$-partite state, $n$-partite
mutual information is ill defined [4].

In order to overcome the problem, in this section, we introduce two definitions about $n$-partite quantum mutual information based on $M_{n}$. By calculating eigenvalues of $M_{n}$, $n$-partite quantum mutual information is defined as the first definition. The second definition is the mathematical expectation of all $2$-partite quantum mutual information in $M_{n}$, as for the Shannon entropy. It is nonnegative and could be easily calculated, and enables us to measure the total correlation on more than two partite. In the end, these definitions are tested and the second definition is proved to be effective.

\emph{Definition 1}. Multipartite quantum mutual information $I'_{G}$ of $n$-partite state $\rho_{A_{1}A_{2}\cdot\cdot\cdot A_{n}}$ can be expressed
\begin{equation}       
I'_{G}(\rho_{A_{1}:A_{2}:\cdot\cdot\cdot: A_{n}})=-\sum_{i}\lambda_{i}\log_{2}\lambda_{i},
\end{equation}
where $\lambda_{i}$ are the eigenvalues of $M_{n}$, and $0\log_{2}0=0$.

From Eq. (5), we know $I'_{G}$ may be negative. So $I'_{G}$ is still an ill definition, and it cannot be selected as the \emph{genuine} quantum mutual information.

\emph{Definition 2}. \emph{Genuine} quantum mutual information $I_{G}$ of $n$-partite state $\rho_{A_{1}A_{2}\cdot\cdot\cdot A_{n}}$ can be expressed

~~~~~~~~~~~~$I_{G}(\rho_{A_{1}:A_{2}:\cdot\cdot\cdot: A_{n}})$
\begin{equation}       
=-\sum_{ij}p(I(\rho_{A_{i}A_{j}}))\log_{2}p(I(\rho_{A_{i}A_{j}})),
\end{equation}
where $p(I(\rho_{A_{i}A_{j}}))=I(\rho_{A_{i}A_{j}})/\sum_{ij}I(\rho_{A_{i}A_{j}})$.

To get some feeling for how the \emph{genuine} quantum mutual information behaves, we now give some properties of it.

\emph{Theorem 3.} (Basic properties of the \emph{genuine} quantum mutual information $I_{G}$)

a) $I_{G}$ is symmetric. It does not change under any permutation of the partite.

b) $I_{G}$ is non-negtive. It is zero if and only if the $n$-partite has the form of $\rho_{A_{1}A_{2}\cdot\cdot\cdot A_{n}}=\rho_{A_{1}}\otimes \rho_{A_{2}}\otimes \cdots \otimes \rho_{A_{n}}$, where $\rho_{A_{i}}$ is a reduced
state of the $i$th subsystem and is pure.

c) $I_{G}(\rho_{A_{1}:A_{2}:\cdot\cdot\cdot: A_{n}})\leq 2\log_{2}n$, with equality if and only if $\rho_{A_{i}}$ is pure with the knowledge of $\rho_{A_{j}}$. Here, $i,j\in \{1,2,\cdots,n\}$.

\emph{Proof}. a) Obvious from the relevant definitions.

b) $-\log_{2}p(I(\rho_{A_{i}A_{j}}))\geq 0$, so $I_{G}\geq 0$ with equality if and only if $I(\rho_{A_{i}A_{j}})=0$. So $S(\rho_{A_{i}})=0$ and $\rho_{A_{i}A_{j}}=\rho_{A_{i}}\otimes \rho_{A_{j}}$, i.e., $\rho_{A_{1}A_{2}\cdot\cdot\cdot A_{n}}=\rho_{A_{1}}\otimes \rho_{A_{2}}\otimes \cdots \otimes \rho_{A_{n}}$ and $\rho_{A_{i}}$ is pure.

c) A very useful property in information theory is $x\log_{2}x$ is a convex function. We find that

~~~~$I_{G}(\rho_{A_{1}:A_{2}:\cdot\cdot\cdot: A_{n}})\leq-\log_{2}\sum_{ij}p^{2}(I(\rho_{A_{i}A_{j}}))$

~~~~~$=-\log_{2}\sum_{ij}I^{2}(\rho_{A_{i}A_{j}})/[\sum_{ij}I(\rho_{A_{i}A_{j}})]^{2}$

~~~~$\leq-2\log_{2}n$.

Notice that equality is achieved if and only if

~~~~~~$I(\rho_{A_{i}A_{j}})=S(\rho_{A_{k}})$, for $i,j,k\in \{1,2,\cdots,n\}$,\\
i.e. $\rho_{A_{i}}$ is pure with the knowledge of $\rho_{A_{j}}$.

Multi-systems are correlated if together they contain more
information than taken separately. If we measure the lack
of information by entropy, this definition of correlations is
captured by the mutual information [10]. The total correlation, as given by the quantum mutual information in Eq. (2), cannot be exhausted by classical correlations and entanglement. Nowadays, there are many ways of understanding the gap in correlations. In the multipartite case it is known that there are several inequivalent classes of states, such as those represented by the $W$-state and the $GHZ$-state. D'Hondt and Panangaden [11] shew that the $W$-state is the only pure state that can be used to exactly solve the problem of leader election in anonymous quantum
networks, and the $GHZ$-state is the only one that can be used to
solve the problem of distributed consensus when no classical post-processing is considered.

In order to gain intuition for the meaning of $I'_{G}$, $I_{G}$, the tripartite information [6,7], and the quantum correlation [12], we consider a 3-qubit $GHZ$-state and a 4-qubit $W$-state respectively. Here, the tripartite information is defined as

$I_{3}(\rho_{A_{1}:A_{2}:A_{3}})=I(\rho_{A_{1}:A_{2}})+I(\rho_{A_{1}:A_{3}})-I(\rho_{A_{1}:A_{2}A_{3}})$;\\
the quantum correlation is the following difference\\
$Q(\rho_{A_{1}A_{2}A_{3}A_{4}})
=\min_{\Pi}[I(\rho_{A_{1}A_{2}A_{3}A_{4}})-I(\Pi(\rho_{A_{1}A_{2}A_{3}A_{4}}))],$
\\where $\Pi(\rho_{A_{1}A_{2}A_{3}A_{4}})=\sum_{\overrightarrow{k}}\Pi^{\overrightarrow{k}}\rho_{A_{1}A_{2}A_{3}A_{4} }\Pi^{\overrightarrow{k}}$, \\ $\overrightarrow{k}=(i_{1},i_{2},i_{3},i_{4})$ and $\Pi_{\overrightarrow{k}}=\Pi_{A_{1}}^{i_{1}}\otimes \Pi_{A_{2}}^{i_{2}}\otimes \Pi_{A_{3}}^{i_{3}}\otimes \Pi_{A_{4}}^{i_{4}}$.

\emph{Example 2}. Consider the following tripartite pure state:
\begin{equation}       
\rho_{A_{1}A_{2}A_{3}}=|GHZ\rangle\langle GHZ|.
\end{equation}
where $|GHZ\rangle=\frac{1}{\sqrt{2}}(|000\rangle+|111\rangle)$.
It has $I_{3}(\rho_{A_{1}:A_{2}:A_{3}})=0$ because the correlations between $A_{1}$ and $A_{2}$ are redundant with those between $A_{1}$ and $A_{3}$. The corresponding quantum mutual information matrix is

~~~~~~~~~~~~~~~~$       
M_{3}=\left(                 
  \begin{array}{ccc}   
    1 & 1& 1 \\  
    1 & 1 & 1\\
    1 & 1 & 1\\
  \end{array}
\right).                 
$\\
Therefore, $I'_{G}(\rho_{A_{1}A_{2}A_{3}})=0$ because the three eigenvalues of $M_{3}$ are respectively 1, 0, 0. $I_{G}(\rho_{A_{1}A_{2}A_{3}})=log_{2}3$, which comes from $I(\rho_{A_{i}A_{j}})=1$ for $i, j\in \{1,2,3\}$, and $p(I(\rho_{A_{i}A_{j}}))=\frac{1}{9}$.

For the maximally entangled state $\rho_{A_{1}A_{2}A_{3}}$, there should exist nonzero quantum correlation because it is not a seperate state. So it has nonzero total correlation, and $I'_{G}(\rho_{A_{1}A_{2}A_{3}})=0$ and $I_{3}(\rho_{A_{1}:A_{2}:A_{3}})=0$ are all not good measurements. On the other hand,

~~~~~~~~~~$I_{G}(\rho_{A_{1}A_{2}A_{3}})=2log_{2}3$,\\
which is just the maximum of $I_{G}(\rho_{A_{1}:A_{2}:A_{3}})$. Therefore, $I_{G}$ in Eq. (7) can be defined as the \emph{genuine} quantum mutual information.

\emph{Example 3}. Consider the following four-partite pure state:
\begin{equation}       
\rho_{A_{1}A_{2}A_{3}A_{4}}=|W\rangle\langle W|,
\end{equation}
where $|W\rangle=\frac{1}{2}(|1000\rangle+|0100\rangle+|0010\rangle+|0001\rangle)$. It has $Q(\rho_{A_{1}A_{2}A_{3}A_{4}})=2$ from Fig.2 in Ref. [12]. So $I(\rho_{A_{1}A_{2}A_{3}A_{4}})\geq 2$.

The reduced density operators are

~~~~~~~~$\rho_{A_{i}}=\frac{1}{4}(3|0\rangle\langle0|+|1\rangle\langle1|,$ \\and

~~~~~~~~$\rho_{A_{i}A_{j}}=\frac{1}{4}(2|00\rangle\langle00|+(|01\rangle+|10\rangle)(\langle01|+\langle10|)$,\\
where $i,j\in \{1,2,3,4\}$ and $i\neq j$. Then, its quantum mutual information matrix is

~~~~~~~~$       
M_{4}=\left(                 
  \begin{array}{cccc}   
    0.8113 & 0.6226& 0.6226&0.6226 \\  
    0.6226 & 0.8113& 0.6226&0.6226\\
    0.6226 & 0.6226& 0.8113&0.6226\\
    0.6226 & 0.6226& 0.6226&0.8113\\
  \end{array}
\right).                 
$\\
The contract diagonal matrix of $M_{4}$ have the four  eigenvalues: 0.2706, 0.3335, 0.6228 and 0.8113. Therefore, $I'_{G}(\rho_{A_{1}A_{2}A_{3}A_{4}})=1.7810$. Through simple calculations, we obtain $I_{G}(\rho_{A_{1}A_{2}A_{3}A_{4}})=3.9897$, which is closer to the maximum value $2\log2(4)=4$ of $I_{G}(\rho_{A_{1}A_{2}A_{3}A_{4}})$.

$I_{G}$ in Eq. (7) is proved as a better multipartite total correlation measurement again.
\section{Conclusion}

The mutual information matrix is $n$ by $n$ real symmetric matrix, and is proved to be always positive semi-definite for all three-tuples [5]. By analogy with it, we define the quantum mutual information matrix which is also $n$ by $n$ real symmetric matrix. However, it is not always positive semidefinite. In this work, we give the necessary and sufficient conditions for the saturating of positive semi-definite characteristic. Further, we have shown that the quantum mutual information matrix can be used to provide a useful tool for characterizing the total correlation in multipartite systems, overcoming some flaws of the $I$-measure [7].

\section*{Acknowledgment}

The authors would like to thank the anonymous reviewers
for helpful comments.

\begin{IEEEbiographynophoto}{Feng Liu}
was born in Heze, China, on June 8, 1980. He received the B.S. degree in applied mathematics from the Liaocheng University, Liaocheng, China, in 2003. And he received the M.S. degree in mathematics from Shannxi Normal University, Xi'an, in 2006. He joined the School of Mathematics and Statistics Science, Ludong University, Yantai, where he participated in the research of cryptography and information theory. He is currently pursuing the Doctorate degree in cryptography at the Beijing University of Posts and Telecommunications (BUPT), where he participates in
the research of quantum cryptography and quantum information theory.
\end{IEEEbiographynophoto}
\begin{IEEEbiographynophoto}{Fei Gao}
was born in Shijiazhuang, China, on January 23, 1980. He received the B.E. degree in communication engineering and the Ph.D. degree in cryptography from the Beijing University of Posts
and Telecommunications, Beijing, China, in 2002 and 2007, respectively. He joined the Network Security Center, State Key
Laboratory of Networking and Switching Technology, Beijing, where he participated in the research of quantum cryptography and quantum information. He is currently working on quantum cryptographic protocols and quantum information theory.

Prof. Gao is a member of the Chinese Association for Cryptologic
Research.
\end{IEEEbiographynophoto}
\begin{IEEEbiographynophoto}{Su-Juan Qin}
was born in Shijiazhuang, China, on August 8, 1979. She received the Ph.D degree in cryptography from the University of Posts and Telecommunications, Beijing, China, in 2008. She is a associate professor with the Network Security Research Center, State Key Laboratory of Networking and Switching Technology, Beijing, China. Her current research interests include quantum cryptographic protocols and quantum information theory.
\end{IEEEbiographynophoto}
\begin{IEEEbiographynophoto}{Qiao-Yan Wen}
was born in Xi'an, China, on July 27, 1959. She received the B.S. and M.S. degrees in mathematics from Shannxi Normal University, Xi'an, in 1981 and 1984, respectively, and the Ph.D. degree in cryptography from Xidian University, Xi'an, in
1997. She is currently a Professor with the Beijing University
of Posts and Telecommunications, Beijing, China, and the Leader of the Network Security Center, State Key Laboratory of Networking and Switching Technology, Beijing. Her current research interests include cryptography, information security, internet security, and applied mathematics.

Prof. Wen is a Senior Member of the Chinese Association for Cryptologic Research.
\end{IEEEbiographynophoto}

\end{document}